\documentclass[twocolumn,showpacs,preprintnumbers,amsmath,amssymb]{revtex4}

\usepackage{graphicx}
\usepackage{dcolumn}
\usepackage{bm}

\begin{document}


\title{Survival of short-range order in the Ising model on negatively curved surfaces}

\author{Yasunori Sakaniwa and Hiroyuki Shima}
\affiliation{Department of Applied Physics, Graduate school of Engineering, Hokkaido University, Sapporo 060-8628, Japan}
\email{shima@eng.hokudai.ac.jp}

\date{\today}

\begin{abstract}
We examine the ordering behavior of the ferromagnetic Ising lattice model 
defined on  a surface with a constant negative curvature. 
Small-sized ferromagnetic domains are observed to exist at temperatures 
far greater than the critical temperature, at which the inner core region 
of the lattice undergoes a mean-field phase transition. 
The survival of short-range order at such high temperatures can be 
attributed to strong boundary-spin contributions to the ordering mechanism, 
as a result of which boundary effects remain active even within 
the thermodynamic limit. 
Our results are consistent with the previous finding of disorder-free 
Griffiths phase that is stable at temperatures lower than the mean-field 
critical temperature.
\end{abstract}



\maketitle

\section{Introduction}

The two-dimensional Ising lattice model is one of the simplest models of 
second-order phase transitions~\cite{Onsager,Kaufman}. 
This model has long played a fundamental role in statistical physics
due its broad applicability and the availability of analytic solutions.
Previous studies have proven that the critical behavior of this model is universal to a large extent, depending on the essential symmetries inherent to the system~\cite{Fisher}.

The two-dimensional Ising lattice model is usually assigned to a flat plane. 
In the last two decades, however, there has been a growing interest 
in the nature of the Ising model assigned to curved 
surfaces~\cite{Rietman,Diego,Hoelbling,Gonzalez,Weigel,Auriac,Deng,Costa,Doyon}. 
This interest is because of its relevance to quantum gravity theory~\cite{Kazakov,Francesco,Holm}
and the successful fabrication of magnetic nanostructures with 
curved geometries~\cite{Yoshikawa,Ji,Liang,Srivastava,Guo,Cabot}. 
In general, the finite curvature of the underlying geometry may alter 
the geometric symmetries of the Hamiltonian describing the system; 
this alteration results in a qualitative change in the critical properties of the system.
In fact, it has been reported \cite{Shima,Shima2,Ueda}
that such curvature-induced alterations occur when the Ising lattice is 
assigned to a surface with a constant negative curvature~\cite{Coxeter,Firdy}. 
Significant shifts in static and dynamic critical exponents toward 
the mean-field values were unveiled in Refs.~\cite{Shima,Shima2},
and the mean-field property of the system was analytically proven 
by employing the renormalization group method~\cite{Ueda}.

The abovementioned findings led to the exploration of surface curvature effects
in various kinds of statistical lattice models~\cite{Milagre,
Hasegawa,Sakaniwa,Hasegawa2,Baek1,Belo,Moura,Krcmar,Gendiar,Krcmar2,Carvalho,Baek2,Baek3,Baek4,Baek5}.
For instance, the $q$-state Potts model on the negatively curved surface exhibited 
a first-order phase transition when $q \ge 3$ \cite{Gendiar},
and the XY model on the same surface showed the absence of 
Kosterlitz-Thouless transition due to strong spin-wave fluctuation~\cite{Baek1}.
It is noteworthy that the concept of negatively curved surfaces (or spaces) is 
also relevant in many fields where the geometric character underlying the system is 
of great significance, such as glass science~\cite{Sausset,Modes,Modes2,Sausset2,Sausset3,Haro,Sausset4},
plasma physics~\cite{Tallez1,Tallez2,Tallez3,Tallez4},
quantum transport~\cite{quant1,quant2},
chaos~\cite{Balazs,Avron,Oloumi,Horwitz}, string theory~\cite{D'Hoker},
and cosmology~\cite{Levin,McInnes}.

Recently, Baek {\it et al.}~\cite{Baek3} studied the percolation transition 
in negatively curved lattices. 
They found two distinct percolation thresholds -- 
one that corresponds to the occurrence of a single infinite-sized cluster 
and the other that corresponds to the occurrence of 
{\it many} infinite-sized clusters connecting a site deeply inside the lattice
to that lying at the outmost boundary of the lattice.
The latter threshold originates from an exponential increase in the total number of sites,
$N\propto e^{L}$, with the linear dimension of the lattice, $L$ (for the definition of $L$, 
see Section II in the present article).
The exponential increase in $N$ leads to an important consequence: 
the ratio of the perimeter to the area of the lattice remains finite 
even within the thermodynamic limit $(L\to \infty)$.
This result is in contrast with the case to that in the case of a flat plane, 
wherein $N \propto L^2$ , {\it i.e.}, the ratio becomes zero for large values of $L$.
Similar percolation thresholds have been observed in enhanced binary trees~\cite{Nogawa}
whose lattice structure is quite analogous
to that considered in Ref.~\cite{Baek3}.
It is conjectured that the two-stage percolation transitions indicate 
the non-uniform ordering of the corresponding Ising model; 
that is, non-vanishing boundary effects even for large $N$
may cause spatially non-uniform growth of ferromagnetic Ising domains 
under cooling, wherein the ordering process near the outmost boundary 
differs from that deep within the lattice. 
The precursor of such non-uniform growth was discussed in Ref.~\cite{Auriac};
the stable Griffiths phase was found near the outmost boundary 
at a temperature lower than the mean-field transition temperature.

In the present article, we examine the ordering process of 
the negatively curved Ising model at temperatures close to and 
far greater than the mean-field transition temperature 
in order to clarify the boundary-spin contributions to 
the domain growth and domain distribution during cooling. 
Using Monte Carlo simulations, we have proved that short-range order 
exists at temperatures far greater than the transition temperature. 
The existence of short-range order at such high temperatures is 
peculiar to the Ising model of negatively curved surfaces and 
is thus an important consequence of non-zero boundary effects 
that are observed even in large systems.

\begin{figure}[ttt]
\includegraphics[scale=0.22]{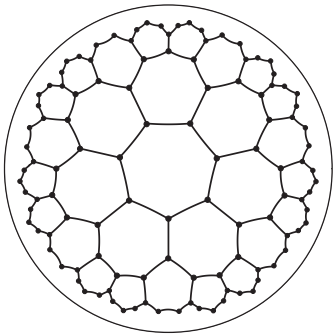}
\caption{Regular heptagonal lattice established on the Poincar\'e disk. 
Here, the number of concentric layers of heptagons $L$ is $3$, and the total 
number of sites is $112$. All heptagons depicted within the 
circle are congruent with respect to the metric given in Eq.~(\ref{eq_03}).}
\label{fig_01}
\end{figure}

\section{Regular lattice on a negatively curved surface}

This section gives a brief summary on the construction of 
regular lattices on a surface with negative curvature. 
A surface with constant negative curvature can be defined as 
a single sheet of a two-sheeted hyperboloid expressed by 
%
\begin{equation}
  x^2+y^2-z^2 = -1 \ \ \ \ (z \ge 1),
\end{equation}
%
which is constructed in the Minkowski space endowed with 
the Minkowskian metric $ds^{2}=dx^{2}+dy^{2}-dz^{2}$.
Although this definition is exact, it is inconvenient for 
computations since three coordinates are used to describe 
a geometry that has only two degrees of freedom. 
We thus use an alternative representation of the surface -- 
the Poincar\'e disk representation -- 
that is obtained by projecting the upper hyperboloid sheet 
onto the $x$-$y$ plane by using the following mapping:
%
\begin{equation}
  (x,y,z) \to \Bigl( \frac{x}{1+z},\frac{y}{1+z} \Bigl).
\end{equation}
%
As a result of this mapping, the upper hyperbolic sheet is transformed into
a unit circle on the {\itshape x-y} plane endowed with the metric 
%
\begin{equation}
ds^{2}=f(dx^{2}+dy^{2}),\ \ f=\frac{4}{(1-x^{2}-y^{2})^{2}}.
\label{eq_03}
\end{equation}
%
This unit circle is referred to as a Poincar\'e disk 
and serves as a convenient representation of the surface 
with a constant negative curvature. 
The Gaussian curvature $\kappa$ on the disk, which is obtained by~\cite{Shima2}
%
\begin{equation}
\kappa=-\frac{1}{f} \biggl( \frac{\partial^{2}}{\partial x^{2}} + 
\frac{\partial^{2}}{\partial y^{2}} \biggl)\log f, 
\label{eq_04}
\end{equation}
%
%
takes the value of $\kappa = -1$ at arbitrary points on the disk.
The boundary of the disk corresponds to points at infinity in the hyperbolic plane.

\begin{figure}[ttt]
\includegraphics[scale=0.42]{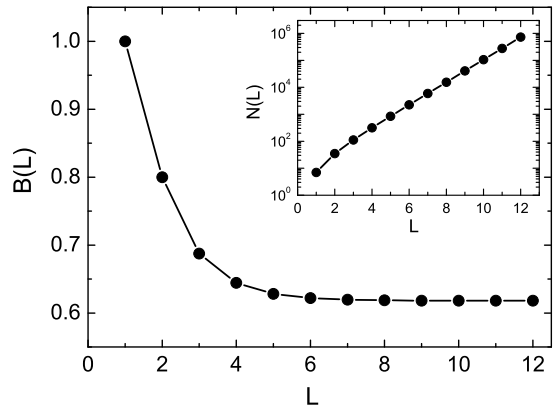}
\caption{Graphical representation of the ratio of 
the number of outmost boundary sites to the number of total sites, 
as expressed by  $B(L)\equiv [N(L)-N(L-1)]/N(L)$.
The horizontal axis represents the number of concentric layers of heptagons $L$. 
Inset: Semi-logarithmic plot of $N(L)$ vs. $L$.}
\label{fig_02}
\end{figure}

An infinite variety of regular polygonal lattices can be built 
on the Poincar\'e disk. All the lattices satisfy the relation $(p-2)(q-2) > 4$, 
where $p$ is the number of edges of each polygon,
and $q$ is the number of polygons around each vertex~\cite{Firdy}. 
The heptagonal lattice of $\{p,q\}=\{7,3\}$ is considered in the present work.
Figure \ref{fig_01} shows the heptagonal lattice represented as a Poincar\'e disk. 
Although polygons depicted in the figure appear to be distorted, 
all of them are exactly congruent with the metric given in Eq.~(\ref{eq_03}).
The size of the entire lattice is characterized by the number of 
concentric layers of heptagons, denoted by $L$, 
which effectively serves as the linear dimension of the lattice.

\begin{figure*}[ttt]
\includegraphics[scale=0.8]{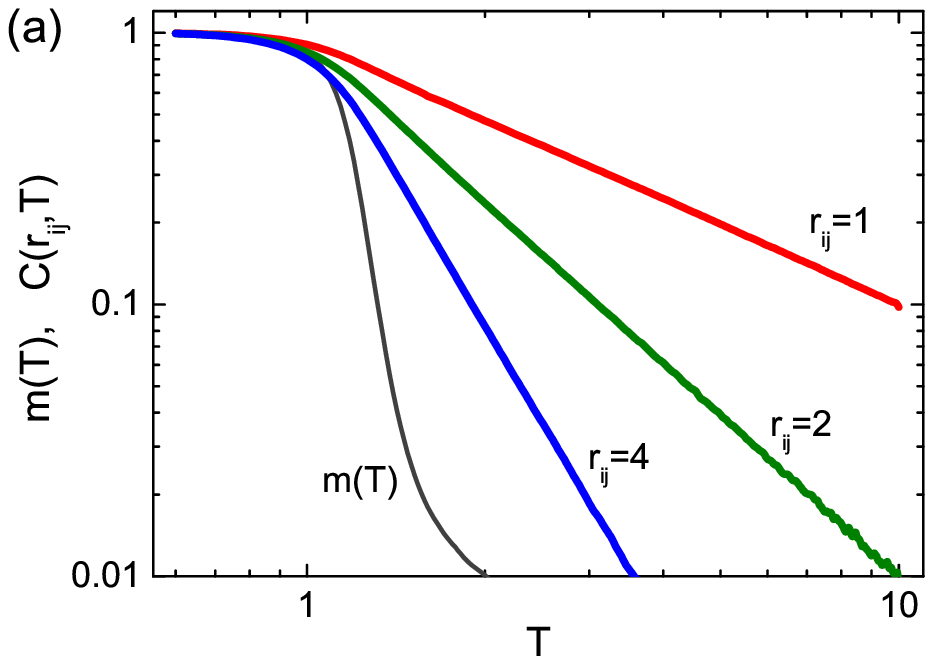}
\includegraphics[scale=0.8]{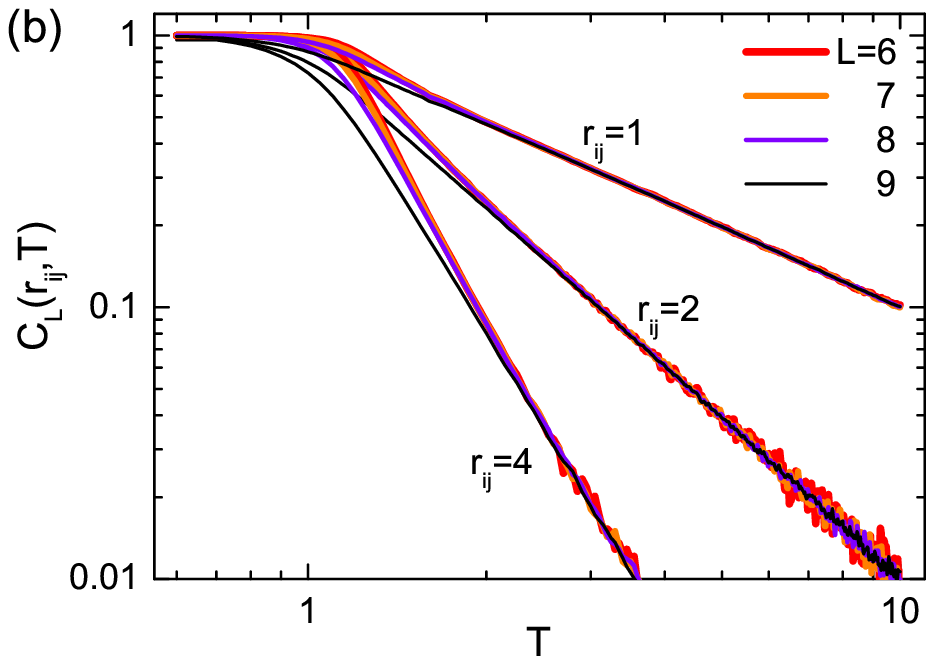}
\caption{(Color online) (a) Double logarithmic plot of spontaneous magnetization $m(T)$ 
(thin line), and two-spin correlation function $C(r_{ij})$ (thick lines).
(b) Correlation functions of two spins both of which lie within
an $L$th concentric layer.}
\label{fig_03}
\end{figure*}

For a given $L$, the total number of sites {\itshape N} is expressed by
%
%
\begin{eqnarray}
N(L) = \left\{
\begin{array}{rl}
&7 \ \ \ \ \mbox{for $L=1$,} \\ [2mm]
&7 + 7 \displaystyle \sum_{\ell=0}^{L-2} 
\left[
u_+ \left( v_{+} \right)^{\ell} + u_- \left( v_{-} \right)^{\ell}  
\right] \ \ \mbox{for $L\ge2$}, 
   \end{array}
  \right.
\label{eq_05}
\end{eqnarray}
%
%
%
where $u_{\pm}=2\pm \sqrt{5}$ and $v_{\pm}=(3 \pm \sqrt5)/2$;
refer to Appendix for the derivation of Eq.~(\ref{eq_05}).
The inset of Fig.~\ref{fig_02} shows the semi-logarithmic plot of $N(L)$ with $L$.
The plot shows that $N$ increases exponentially for $L \gg 1$.
In fact, we obtain
%
\begin{equation}
B(L) \equiv \frac{N(L)-N(L-1)}{N(L)} \to 1-\frac{1}{v_+}, \ \ \ (L \to \infty)
\label{eq_06}
\end{equation}
%
which implies the boundary contribution quantified by 
$B(L)$ does not become zero, but remains finite within the limit $L \to \infty$.
It is emphasized that non-vanishing property of $B(L)$ as well as
the exponential increase in $N(L)$ with $L$ 
is a result of the constant negative curvature of the underlying geometry, 
and it is the reason behind the non-trivial critical behavior of 
embedded lattices, as proved in previous studies~\cite{Shima,Shima2}.
%
%

\section{Numerical method}

The current study aims at exploring the ordering mechanism of the heptagonal Ising model
with ferromagnetic interaction.
The Hamiltonian of the heptagonal Ising model is given by 
%
%
\begin{equation}
 \mathcal{H} = -J\sum_{\langle i,j \rangle}s_{i}s_{j},\ \ s_{i}=\pm 1, 
\end{equation}%
%
where $\langle i,j \rangle$ denotes a pair of nearest-neighbor sites 
on a lattice with free boundary conditions. 
The coupling strength $J$ is a constant, considering the fact that 
all the Ising spins are equally-spaced on the Poincar\'e disk. 
Throughout this paper, $J/k_{B}$ and $J$ are used as the units of 
temperature and energy, respectively. 
We have employed Monte Carlo simulations to calculate the 
order parameter $m=\langle s_i \rangle$ 
and the two-point correlation function 
$C(r_{ij}) = \langle s_{i} \cdot s_{j} \rangle$ with $r_{ij} \equiv |\bm{r}_{i}-\bm{r}_{j}|$.
Here, $\bm{r}_i$ represents the location of the $i$th spin on the Poincar\'e disk,
and the distance $r_{ij}$ is a measure of the number of bonds 
along the shortest path between the $i$th and $j$th sites.
Configuration space sampling has been carried out 
by using a cluster-flip algorithm~\cite{Wolff}, 
and averages have been calculated by considering $10^{5}$ samples.

%
\section{Results and discussion}
%
%

Figure \ref{fig_03} shows the temperature dependence of $C(r_{ij},T)$ for different $r_{ij}$'s.
We have set $L=9$ ({\it i.e.}, $N=40432$) in all calculations. 
We found that for $r_{ij} \le 4$, 
$C(r_{ij})$ shows power-law decay with $T$ and thus survive at $T > T_{c} \simeq 1.25 $ \cite{Shima}
though $m(T)$ is almost zero there.
These survival of $C(r_{ij})$ indicates that small-sized ferromagnetic domains 
remain active even at high temperatures, 
although they do not contribute to $m(T)$ because of the change in sign 
between positive ($s_i = +1$) and negative ($s_i = -1$) domains.
To explore the variety of domain sizes and examine boundary effects on size variations, 
we evaluate the correlation function, which is denoted by
$C_{L}(r_{ij},T)$, of two spins that lie in the $L$th layer. 
Figure \ref{fig_03}(b) shows the $T$-dependence of $C_{L}$ for $r_{ij}=1,2,4$, 
in which $L$ is varied from $6$ to $9$
(data for $L\le 5$ are omitted since they are indistinguishable from data for $L=6$).
Interestingly, the data of $C_{L}$ for each $r_{ij}$ value collapses
onto a single curve at $T > T_c$. 
This collapse implies that  the domains that continue to exist at $T > T_c$
are of the same size and are uniformly scattered over the entire lattice, 
regardless of their distance from the outmost boundary.

It is emphasized that the persistence of domains at such high temperatures is 
supplementary to previous findings~\cite{Auriac,Shima} related to 
low-temperature behaviors of an identical system. 
It was found that at $T<T_c$, the central region of 
the lattice (far from the outmost boundary) exhibits an ordered phase 
that is a consequence of a mean-field transition~\cite{Shima},
while the outer region (close to the boundary) exhibits 
a stabilized Griffiths phase that is free from extrinsic disorder~\cite{Auriac}.
Note that the former result indicates the presence of a paramagnetic phase 
in the central region at $T>T_c$, 
where Ising spins are randomly oriented due to large thermal fluctuations.
Hence, one would expect the formation of ferromagnetic domains to be hindered at $T>T_c$;
however, contrary to the expectation, the formation is observed in the current simulations.
This apparent contradiction is attributed to the finite-size effect.
For a moderagely large $L$, the strong boundary-spin contributions penetrate to the center of the lattice,
as a result of which small-sized domains arise not only near the boundary but also in the central region.
We conjecture that the domain in the central region disappears if a sufficiently large value of $L$
is considered: however, huge computational costs are involved when the value of $L$ is large enough.

%
%
\begin{figure}[ttt]
\includegraphics[scale=0.8]{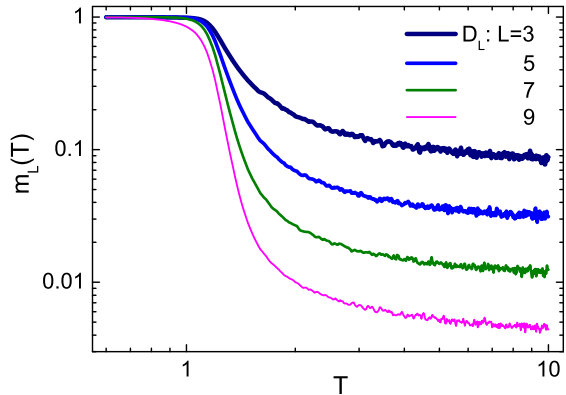}
\caption{(Color online) Spontaneous magnetization $m_L(T)$ within a circular region
${\cal D}_L$ enclosed by the $L$th layer.}
\label{fig_04}
\end{figure}

To gain a better insight into the above issue, we evaluate the quantity
$m_L(T) \equiv \frac{1}{N} \sum s(\mathbf{r}_i \subset {\cal D}_L)$, where 
${\cal D}_L$ is a circular region enclosed by the $L$th layer.
Figure \ref{fig_04} shows the $T$-dependence of $m_L(T)$ for different values of $L$.
It is found that above $T_{c}$, $m_L(T)$ for $L=3$ is finite whereas $m(T)$ becomes almost zero.
This indicates an incomplete cancellation between the positive and negative domains 
in the region ${\cal D}_{L=3}$ due to its small size.
We emphasize that $m_L(T)$ should become zero if ${\cal D}_L$ is 
in the paramagnetic phase with no finite-sized domains.
Therefore, the finiteness of $m_{L=3}(T)$ at $T>T_c$ is another piece of evidence 
that supports the persistence of small-sized domains.
We have also confirmed that finite $m_{L=3}(T)$ at high $T$
tends to disappear with an increase in $L$, since boundary-spin contributions
are prevented from penetrating to the center of the lattice.

\begin{figure}[ttt]
\includegraphics[scale=0.13]{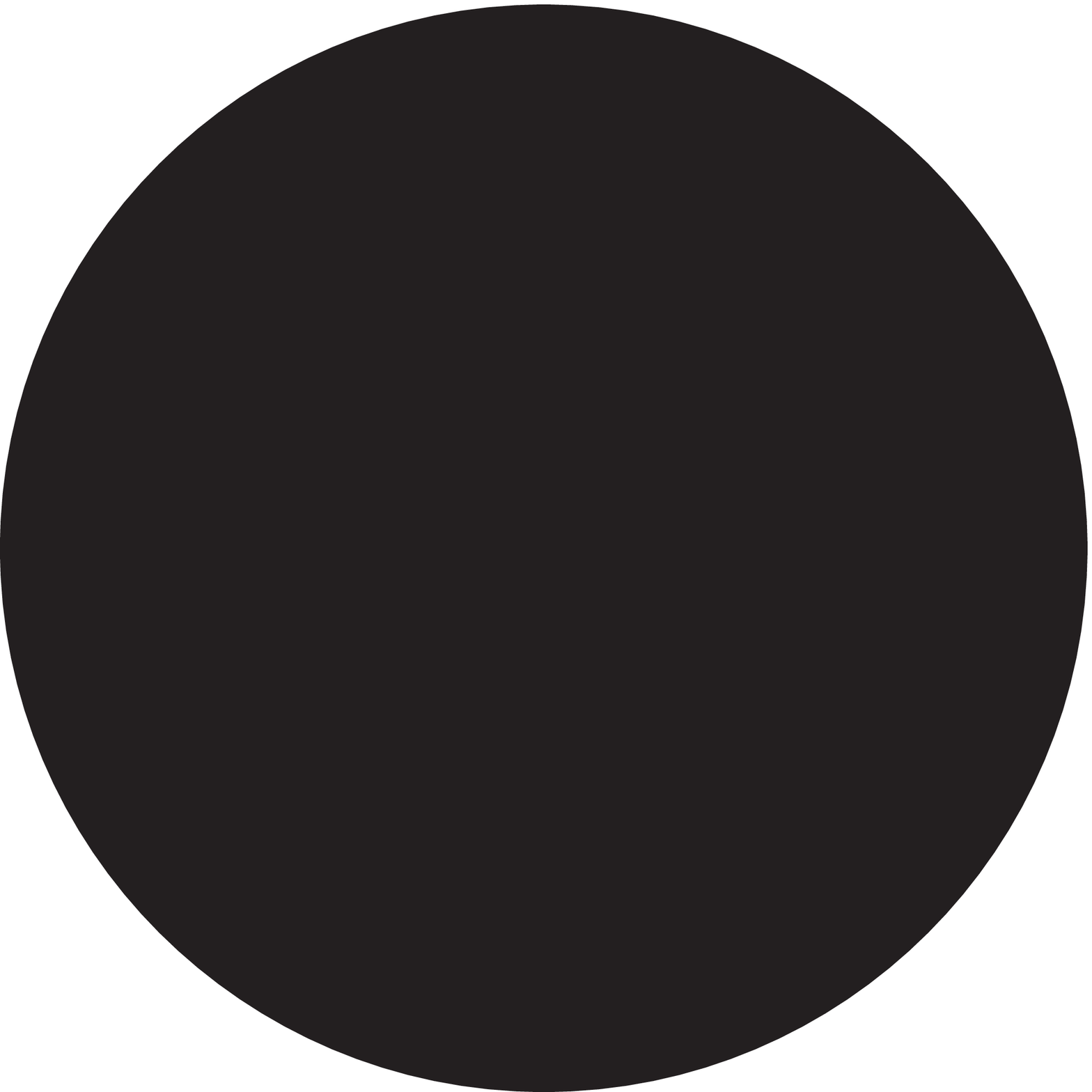}
\includegraphics[scale=0.13]{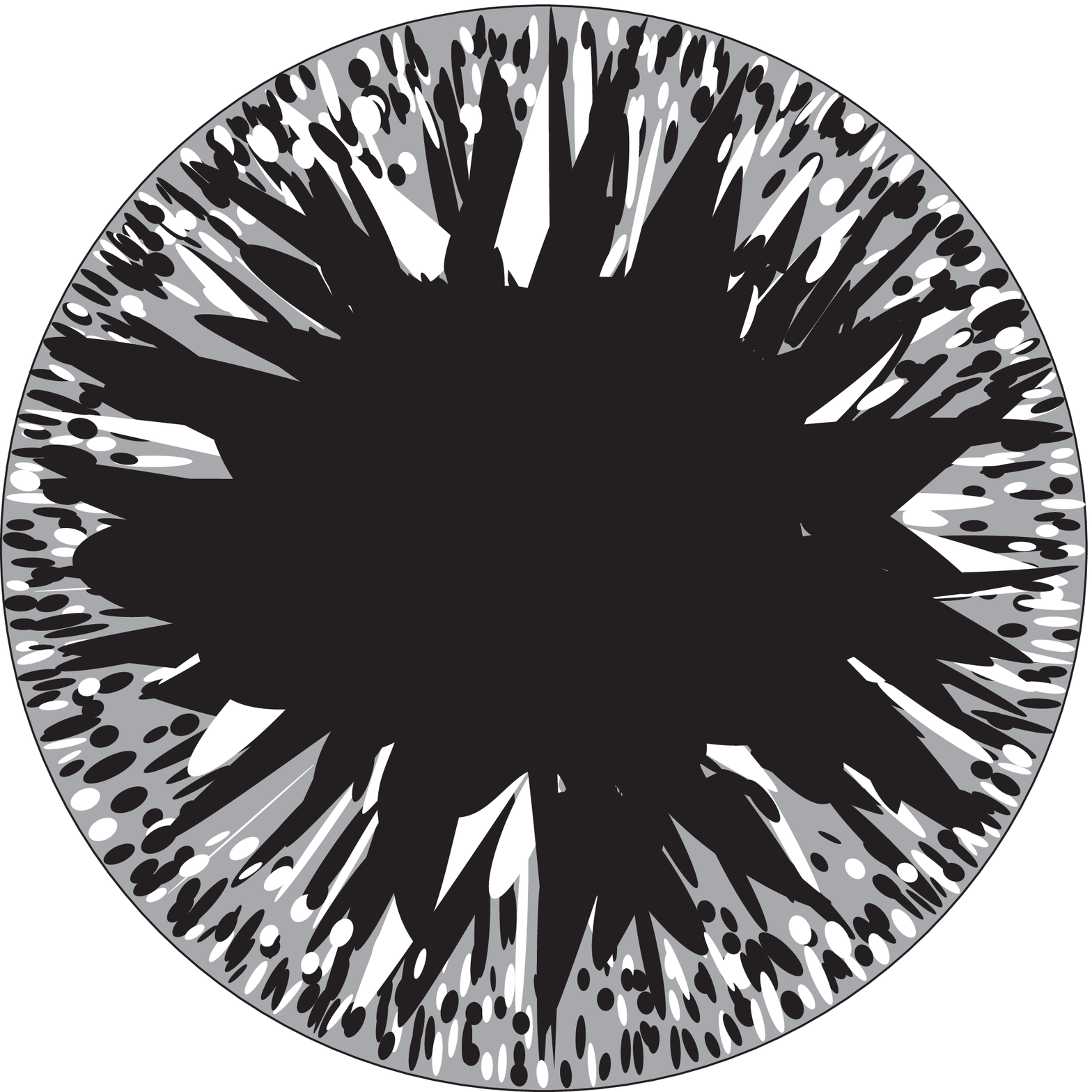}
\includegraphics[scale=0.13]{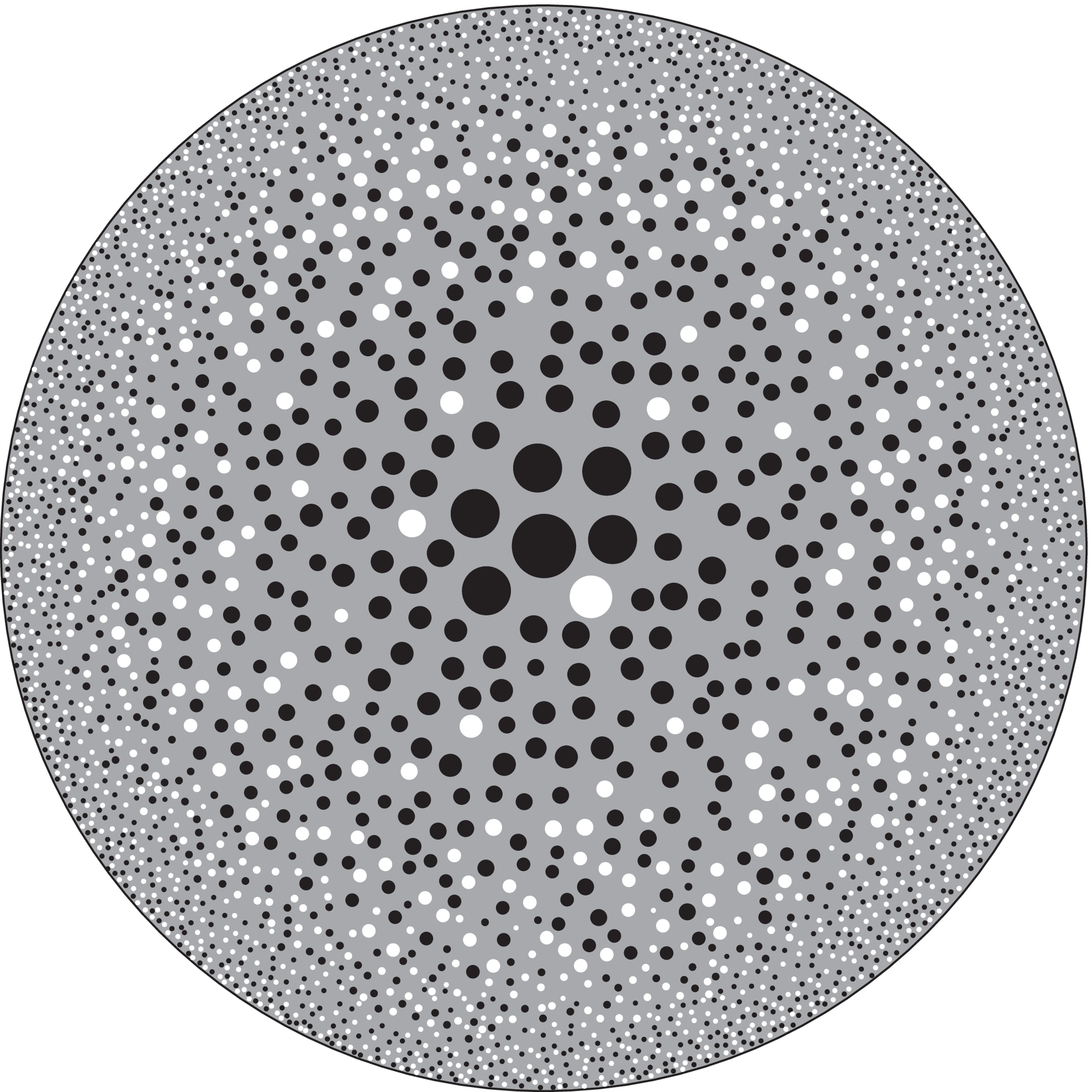}
\caption{Illustration of the ordering process: the left panel represents $T < T_c \simeq 1.25$;
the middle panel, $T \sim T_c$; and the right panel, $T > T_c$.
The open and solid circles indicate ferromagnetic domains with positive and negative directions,
respectively. 
The gray area represents the paramagnetic phase.}
\label{fig_05}
\end{figure}

Figure \ref{fig_05} illustrates the ordering process of our moderately large Ising lattice with 
a decrease in the temperature.
At $T \gg T_c$, small-sized ferromagnetic domains (indicated by open and solid circles) 
are randomly embedded in the paramagnetic phase which is shown in gray. 
These domains are distributed homogeneously across the lattice,
but give rise to finite $m_{L}(T)$ for small values of $L$ due to 
the incomplete cancellation between the positive and negative domains. 
Subsequently, a decrease in $T$ results in the growth of domains within the inner region.
Eventually, the ferromagnetic phase is formed through the mean-field phase transition.
Nevertheless, near the outmost boundary, small domains remain active and fluctuate
on a large time scale, as observed in Ref.~\cite{Auriac}.

\section{Conclusion}

In the present work, we have considered the ordering mechanism of 
the Ising lattice model assigned to a negatively curved surface. 
We have found that small-sized domains survive at temperatures much higher than $T_c$,
at which the inner bulk region of the lattice goes through mean-field phase transition. 
The existence of small domains is attributed to the non-zero boundary-spin 
contribution that is unique to negatively curved surfaces. 
Our results are consistent with those of previous studies: 
the Griffiths phase is present near the outmost boundary of the lattice,
while the inner core region undergoes the mean-field phase transition.
This consistency sheds light on the nature of phase ordering 
in a wide variety of statistical lattice models assigned to 
negatively curved surfaces, most of which need to be further studied.

\section*{Acknowledgements}
The authors are grateful to K.~Yakubo, T.~Hasegawa and T.~Nogawa
for fruitful discussions. Y.S is thankful for the financial 
support from JSPS Research Fellowships for Young Scientists. 
H.S acknowledges the support from the Executive Office for 
Research Strategy in Hokkaido University. 
This work was supported in part by Grant-in-Aid for Scientific Research from Japan Ministry of 
Education, Science, Sport and Culture.
Numerical calculations were performed in part by facility at the Supercomputer Center, 
ISSP, University of Tokyo.

\appendix

\section{Proof of the formula (5)}

\begin{figure*}[ttt]
\includegraphics[scale=0.25]{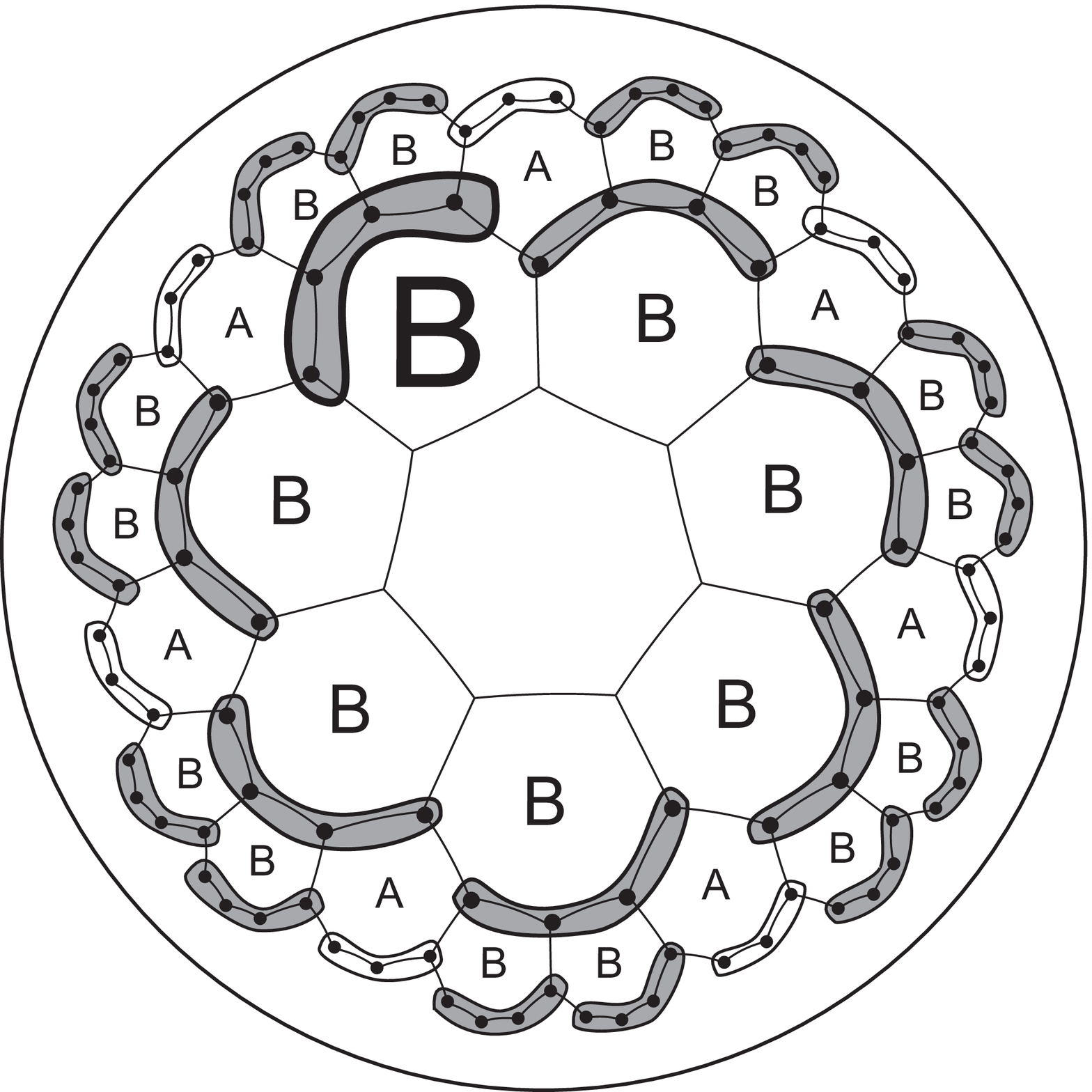}
\hspace*{36pt}
\includegraphics[scale=0.35]{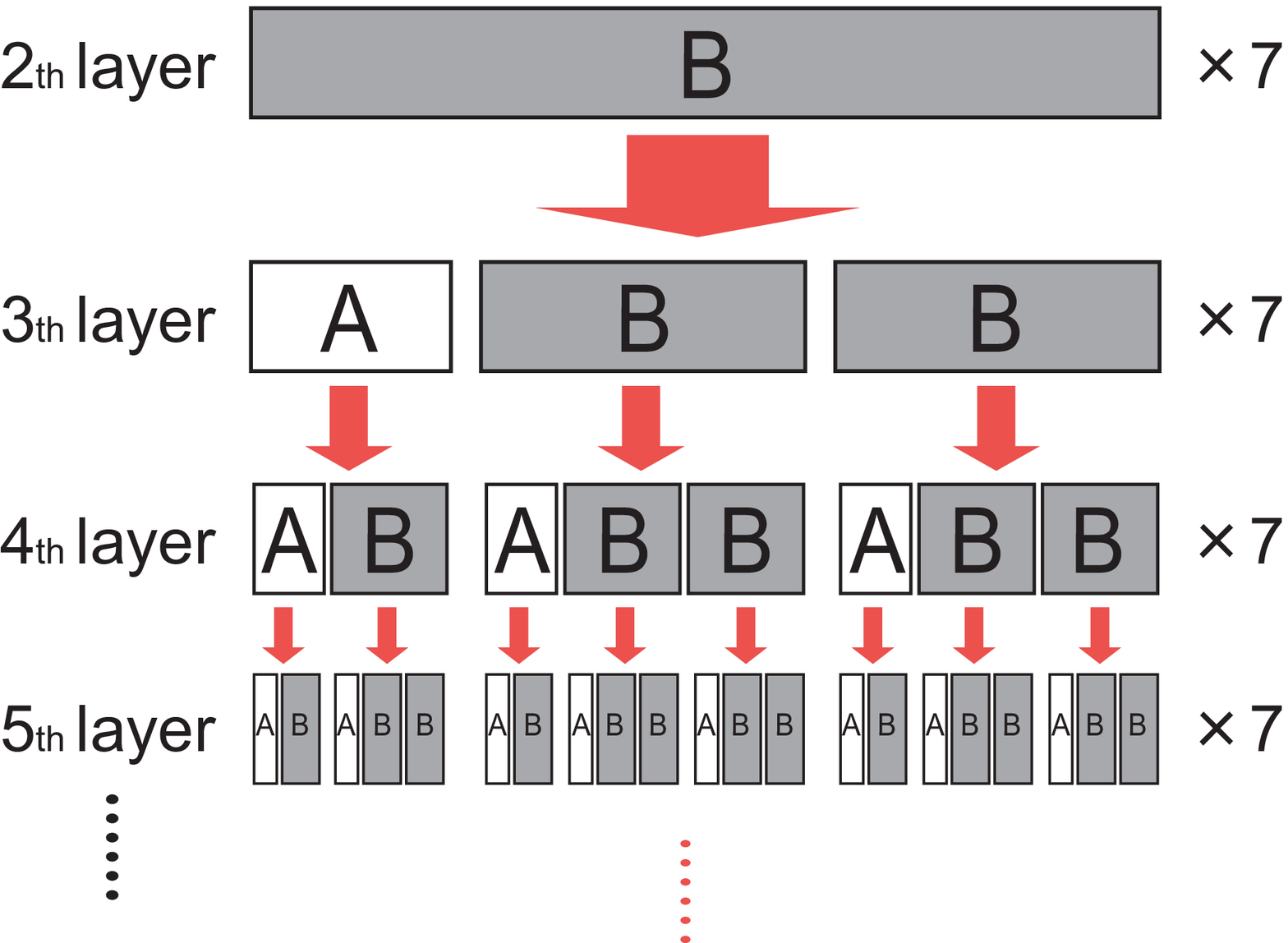}
\caption{(Color online) (a) Classification scheme of heptagons and associated sites in the lattice. 
All heptagons are classified into two groups as indicated by symbols $A$ and $B$ 
(see text for the method of classification). A group of four (or three) sites encircled by a 
curve is attributed to those of the attached $B$- (or $A$-) heptagon; 
the four sites grouped by a very thick curve, for instance, are regarded as those of the heptagon 
marked by the bold ``$B$'' index. (b) Proliferation diagram of $A$ and 
$B$ heptagons in the lattice. Each $A$ and $B$ block in the diagram 
symbolizes the associated group of three and four sites, respectively, at the $L$th layer.}
\label{growing_rule}
\label{fig_app01}
\end{figure*}

This Appendix is devoted to the derivation of Eq.~(\ref{eq_05})
--- the explicit form of $N(L)$ as a function of $L$. 
We classify all the congruent heptagons in the lattice into two groups: 
(i) heptagons having a common edge with the adjacent heptagon lying 
in the inner concentric layer (referred to as ``$A$"-heptagons)
and (ii) those having two common edges with the adjacent one heptagon in the inner layer 
(referred to as ``$B$''-heptagons). 
Figure \ref{fig_app01}(a) illustrates our classification, 
in which either of the symbols $A$ or $B$ is assigned to all heptagons 
except for the central one.

We see in Fig.~\ref{fig_app01} that the second innermost layer $(L=2)$ consists 
of seven $B$'s, the third innermost layer $(L=3)$ consists of 14 $B$'s and seven $A$'s, and so on.
We identify the four sites within a hatched region as those attached to the 
$B$-heptagon; in a similar way, the three sites enclosed by a curve are 
identified as those attached to the $A$-heptagon. 
Then, we can say that the four sites of a $B$-heptagon located at $L=2$ ``engenders'' one $A$-heptagon and 
$B$-heptagons at $L=3$. 
Similarly, it follows that an $A$-heptagon at $L=3$ generates a pair 
containing an $A$-heptagon and a $B$-heptagon at $L=4$ (although it is not shown in 
Fig. \ref{fig_app01}(a)). 
This proliferation process is summarized by the diagram in Fig.~\ref{fig_app01}(a); 
each $A$ (or $B$) block symbolizes an ensemble of three (or four) sites, and a block 
in the $L$th layer creates several blocks in the $(L+1)$ layer. 
The diagram illustrates a method to count the total number of sites contained 
in a given $L$th layer.

We now have all the ingredients to derive the formula.
Let us denote the number of $A$- (or $B$-) heptagons
in the $L$th layer by $a_L$ (or $b_L$) and set $a_2=0, b_2=7$.
It follows from the diagram that for any $L \ge 3$, 
%
%
\begin{equation}
a_{L} = a_{L-1}+b_{L-1}, \;\;\; b_{L} = a_{L-1}+2b_{L-1}.
\label{eq_app01}
\end{equation}
%
%
We eliminate $b_L$ from Eq.~(\ref{eq_app01}) to obtain $a_L - 3a_{L-1} + a_{L-2} = 0$
whose solutions under the conditions $a_2 = 0$ and $a_3 =7$ are given by 
%
%
\begin{equation}
a_L = \frac{7\sqrt5}{5} 
\left[ \bigl( v_+ \bigl)^{L-2} - \ 
\bigl( v_- \bigl)^{L-2} \ 
\right],\ \ \mbox{for $L\ge3$}
\end{equation}
%
%
where $v_{\pm} = (3 \pm \sqrt{5})/2$. 
Hence, we have for $L\ge2$,
%
%
\begin{equation}
b_L = \frac{7\sqrt5}{10} 
\left[
w_+ \bigl( v_{+} \bigl)^{L-2} +
w_- \bigl( v_{-} \bigl)^{L-2}
\right],
\end{equation}
%
%
where $w_{\pm} = \pm 1 + \sqrt{5}$.
As a result, the desired formula -- Eq.~(\ref{eq_05}) -- 
for $N(L) \equiv 3a_L + 4b_L$ can be obtained.


\end{document}